\documentclass[a4paper,11pt]{article}
\pdfoutput=1
\usepackage{jcappub}
\usepackage{aas_macros}
\usepackage{subcaption}

\renewcommand{\d}{\mathrm{d}} 
\renewcommand{\L}{\mathcal{L}} 
\newcommand{\Lbar}{\bar{\mathcal{L}}} 
\newcommand{\vk}{\mathbf{k}}
\newcommand{\vkhat}{\mathbf{\hat{k}}}
\newcommand{\vrhat}{\mathbf{\hat{r}}}
\newcommand{\vnhat}{\mathbf{\hat{n}}}

\newcommand{\vr}{\mathbf{r}}
\newcommand{\vrone}{\mathbf{r}_1}
\newcommand{\vrtwo}{\mathbf{r}_2}

\newcommand{\real}{\textrm{r}}

\newcommand{\Ylm}{Y_{\ell m}}

\newcommand{\lmax}{\ell_{\rm max}}

\newcommand{\Pobs}{P^{\rm obs}}

\newcommand{\dirac}{\delta_{\mathrm{D}}}
\renewcommand{\l}{\ell}
\newcommand{\lprime}{\ell^\prime}
\DeclareMathOperator{\cov}{Cov}
\DeclareMathOperator{\FT}{FT}

\title{An optimal FFT-based anisotropic power spectrum estimator}

\author[a,b]{Nick Hand,}
\author[b,c,d,e]{Yin Li,}
\author[b,c]{Zachary Slepian,}
\author[a,b,c,d]{\& Uro{\v s} Seljak}

\affiliation[a]{Department of Astronomy, University of California, Berkeley, CA 94720, USA}
\affiliation[b]{Lawrence Berkeley National Laboratory, Berkeley, CA 94720, USA}
\affiliation[c]{Berkeley Center for Cosmological Physics, University of California, Berkeley, CA 94720, USA}
\affiliation[d]{Department of Physics, University of California, Berkeley, CA 94720, USA}
\affiliation[e]{Kavli Institute for the Physics and Mathematics of the Universe (WPI),
UTIAS, The University of Tokyo, Chiba 277-8583, Japan}

\emailAdd{nhand@berkeley.edu}
\emailAdd{yin.li@berkeley.edu}
\emailAdd{zslepian@lbl.gov}
\emailAdd{useljak@berkeley.edu}

\abstract{
Measurements of line-of-sight dependent clustering via the galaxy power spectrum's multipole moments constitute a powerful tool for 
testing theoretical models in large-scale structure. Recent work shows that 
this measurement, including a moving line-of-sight, can be accelerated using 
Fast Fourier Transforms (FFTs) by decomposing the Legendre polynomials into 
products of Cartesian vectors.
Here, we present a faster, optimal means of 
using FFTs for this measurement. We avoid redundancy present in the Cartesian 
decomposition by using a spherical harmonic decomposition of the Legendre polynomials.
Consequently, our method is substantially faster: a given multipole of order $\ell$ 
requires only $2\ell+1$ FFTs rather than the $(\ell+1)(\ell+2)/2$ FFTs of 
the Cartesian approach. For the hexadecapole ($\ell = 4$), this translates 
to $40\%$ fewer FFTs, with increased savings for higher $\ell$.
The reduction in wall-clock time enables the calculation of finely-binned wedges 
in $P(k,\mu)$, obtained by computing multipoles up to a large $\ell_{\rm max}$ 
and combining them. This transformation has a number of advantages. 
We demonstrate that by using non-uniform bins in $\mu$, we can isolate 
plane-of-sky (angular) systematics to a narrow bin at $\mu \simeq 0$ while
eliminating the contamination from all other bins. We also show that the 
covariance matrix of clustering wedges binned uniformly in $\mu$ becomes 
ill-conditioned when combining multipoles up to large values of $\lmax$, 
but that the problem can be avoided with non-uniform binning. As an 
example, we present results using $\lmax=16$, for which our 
procedure requires a factor of 3.4 fewer FFTs than the 
Cartesian method, while removing the first $\mu$ bin leads only to a 7\% increase
in statistical error on $f \sigma_8$, as compared to a 
54\% increase with $\lmax=4$.
}

\begin{document}
\maketitle

\section{Introduction}

The clustering of galaxies on the largest scales contains a significant
amount of cosmological information. The Baryon Acoustic Oscillation (BAO) 
feature on scales of $\sim 100\;{\rm Mpc}/h$ can be used as a standard ruler 
to gauge the Universe's expansion history and infer properties of dark energy 
(e.g., \cite{WagnerMullerSteinmetz08,ShojiJeongKomatsu09}).
First detected in the 2-point correlation function (2PCF) by
\cite{EisensteinZehaviEtAl05,ColePercivalEtAl05} and more recently in
the 3-point function (3PCF) by \cite{SlepianEisensteinEtAl16}, the BAO 
signal has provided percent-level measurements 
of the Hubble parameter $H(z)$ and angular diameter distance 
$D_a(z)$ \cite{AlamEtAl16}. These analyses have measured both the 
characteristic BAO peak in configuration space 
\cite{RossEtAl17,Vargas-MaganaEtAl17} and the analogous 
wiggles in Fourier space \cite{BeutlerEtAl17,Gil-MarinEtAl16}. 
Beyond the BAO, and on even larger scales, these clustering statistics
also contain signatures of primordial non-Gaussianity, the deviation 
from Gaussian Random Field initial conditions in the early Universe
\cite{CreminelliNicolisEtAl06,DesjacquesSeljak10b}.

Additional information can be extracted from these statistics by
measuring the broadband clustering as a function of the angle to the 
line-of-sight (LOS). Although the underlying distribution of galaxies is 
assumed to be homogeneous and isotropic, observational effects such as 
the Alcock-Paczynski (AP; \cite{AlcockPaczynski79}) effect and 
redshift-space distortions (RSD; \cite{Kaiser87}) introduce anisotropy 
into the measured clustering when a fiducial distance-redshift relation is 
used to translate redshifts into comoving coordinates. In particular, 
anisotropy around the line-of-sight is introduced by RSD because an object's 
redshift, used to infer the LOS coordinate, 
is sensitive to its peculiar velocity. Because this velocity is sourced by the 
large-scale gravity field, RSD measurements allow constraints on the growth 
rate of structure and can 
provide tests of General Relativity (e.g., \cite{GuzzoPierleoniEtAl08}).
For galaxy pairs, RSD depend on the angle cosine $\mu$ between the pair separation 
$\vec{s}$ and the line-of-sight $\hat{n}$. The clustering is
typically measured as multipole 
moments of the 2-point correlation function, which gives the excess of 
pairs above random, or of the power spectrum, its Fourier space analog. 
The Legendre polynomials form a complete basis and are equivalent to expanding 
in powers of $\mu$. Parity demands that only even multipoles are 
non-zero. In linear theory, RSD generate only $\ell = 0, 2,$ and $4$ 
moments of the anisotropic power spectrum 
\cite{Kaiser87} or correlation function \cite{Hamilton92}.  

For wide-field galaxy surveys, only angle-averaged clustering, i.e., the monopole, 
can be measured accurately under the assumption of a single LOS to the 
entire survey. Under this assumption, it is straightforward to measure the clustering
using a Fast Fourier Transform (FFT). What is more challenging is to define a
clustering estimator for the higher-order multipoles that uses a line-of-sight that 
rotates to follow each galaxy pair's spatial or Fourier-space separation. 
Including the observer as a third vertex, the galaxy pair maps to a triangle, and more accurate 
line-of-sight choices are the angle bisector of this triangle or the vector 
from the observer to the separation midpoint. Less accurate but still better than 
a single LOS is taking the LOS to be the vector from observer to a single pair 
member, as first used in \cite{YamamotoEtAl06,BlakeEtAl11}. This latter method,
often referred to as the local plane-parallel approximation,
differs from angle bisector and midpoint methods at $\mathcal{O}(\theta^2)$, where 
$\theta$ is the angle the pair subtends; bisector and midpoint methods also differ from 
each other at $\mathcal{O}(\theta^2)$ \cite{Slepian15_wide_angle}. For the 
current generation of surveys, these wide-angle effects are not a significant
source of error \cite{SamushiaBranchiniPercival15,YooSeljak15} but could become
important for future surveys, especially for studies which focus primarily on 
large-scales, i.e., primordial non-Gaussianity analyses. To address this, slight generalizations of the local plane parallel estimates for the multipoles can be combined to form the midpoint and bisector-based estimates \cite{Slepian15_wide_angle}.


Recently, \cite{BianchiGil-MarinEtAl15} and \cite{Scoccimarro15} showed that by using 
products of Cartesian coordinates as building blocks for the Legendre polynomials,
one could evaluate the local plane-parallel method of \cite{YamamotoEtAl06} using FFTs,
providing an enormous speed-up over the summation-based estimator. Around the same time,
\cite{SlepianEisenstein16} demonstrated that FFTs could also be used for the 
anisotropic 2PCF by exploiting the spherical harmonic addition theorem to decompose the 
Legendre polynomials into spherical harmonics. In this work, we show that 
this spherical harmonic approach can also be used for the power spectrum multipoles.
Importantly, the spherical harmonics are orthogonal to each other, whereas the 
Cartesian vectors used by \cite{BianchiGil-MarinEtAl15,Scoccimarro15} are not.
Thus, the \cite{BianchiGil-MarinEtAl15,Scoccimarro15} implementation involves redundancy, requiring $(\ell+1)(\ell+2)/2=\mathcal{O}(\ell^2)$ FFTs per multipole rather
than the $2\ell+1=\mathcal{O}(\ell)$ FFTs needed by our method. We emphasize that our algorithm scales linearly with $\ell$ whereas these previous works scaled with its square.

The additional speed-up provided by our implementation is not
only useful for computing higher-order multipoles more quickly but also 
for the processing of a large number of mock catalogs for estimating covariance matrices.
For example, the covariance matrix estimation of \cite{AlamEtAl16} required evaluating clustering statistics for 3 separate redshift bins 
and 1000 mock catalogs. Furthermore, the calculation of higher-order multipoles 
is also useful for analyzing the clustering in wedges of $\mu$ 
\cite{KazinSanchezBlanton12, GriebEtAl16}. While there is little measurable signal in 
multipoles above the $\ell=4$ hexadecapole, we show that 
the measurement of multipoles up to a large $\lmax$ allows the use of
narrow $\mu$ bins. It also reduces the correlations between separate $\mu$ bins,
allowing for easier theoretical modeling of the covariance of the clustering estimator.
The use of narrow $\mu$ wedges becomes advantageous when measuring
clustering contaminated by systematics in the plane of the sky, as is often
the case for galaxy surveys, i.e., \cite{PinolEtAl17}. Such a transverse systematic 
will contaminate all multipoles, but we demonstrate that the
contamination can be effectively isolated to a narrow bin around $\mu \simeq 0$ 
when using wedges, with the width of the $\mu\simeq 0$ bin scaling as 
$(\lmax/2+1)^{-1}$. Non-uniform binning in $\mu$ can be chosen such that any 
artifacts of the systematic are eliminated for all bins beyond the first 
$\mu\simeq 0$ bin. 

The paper is laid out as follows.
In \S\ref{sec:multipoles}, we first present the improved estimator of 
the power spectrum multipoles using a spherical harmonic expansion
and demonstrate that it significantly outperforms the Cartesian decomposition method.
This enables us to efficiently measure higher-order multipoles
and then transform them into power spectrum wedges as shown in \S\ref{sec:wedges}.
We then discuss our implementation of the estimators in the publicly available
large-scale structure analysis software \texttt{nbodykit} in \S\ref{sec:implementation}.
In \S\ref{sec:systematics}, we develop a simple model for a systematic signal
in the transverse ($\mu=0$) direction and present a simple method to 
mitigate the contamination with a non-uniform binning scheme. We discuss the impact
of survey window function on this method in \S\ref{sec:win}.
We show in \S\ref{sec:cov} that the higher multipoles de-correlate the wedges
even though they do not add additional signal. This means that one can reduce the information loss due to removal of the localized contamination
by measuring more multipoles (\S\ref{sec:info}).
Finally, we conclude in \S\ref{sec:conclusions}.

\section{Estimators}\label{sec:estimators}

\subsection{Multipoles}\label{sec:multipoles}

We begin by defining the weighted galaxy density field \cite{FeldmanKaiserEtAl94},

\begin{equation}\label{eq:FKP-density}
F(\vr) = \frac{w(\vr)}{I^{1/2}} \left [n(\vr) - \alpha n_s(\vr) \right],
\end{equation}
where $n$ and $n_s$ are the observed number density field for the galaxy catalog and
synthetic catalog of random objects, respectively. The random catalog defines
the expected mean density of the survey and also accounts for the angular mask and radial 
selection function. It contains no cosmological clustering signal. We allow
for a general weighting scheme $w(\vr)$. The factor $\alpha$ normalizes
the synthetic catalog to the number density of the galaxies. The field
$F(\vr)$ is normalized by the factor of $I$, defined as 
$I \equiv \int d\vr\; w^2 \bar{n}^2(\vr)$. The estimator for the multipole 
moments of the power spectrum is \cite{FeldmanKaiserEtAl94,YamamotoEtAl06}

\begin{equation}\label{eq:Pell-full}
\widehat{P}_\ell(k) = \frac{2\ell+1}{I} \int \frac{d\Omega_k}{4\pi} 
									\left [ 
									\int \d\vrone
                                    \int \d\vrtwo \ F(\vrone) F(\vrtwo) 
                                    e^{i \vk \cdot (\vrone-\vrtwo)}
                                    \L_\ell(\vkhat \cdot \vrhat_h)
                                    - P_\ell^{\rm noise}(\vk)
                                    \right ],
\end{equation}
where $\Omega_k$ represents the solid angle in Fourier space, 
$\vr_h \equiv (\vrone + \vrtwo)/2$ is the line-of-sight to the mid-point
of the pair of objects, and $\L_\ell$ is the Legendre polynomial of order $\ell$.
The shot noise $P_\ell^{\rm noise}$ is

\begin{equation}
P_\ell^{\rm noise}(\vk) = (1 + \alpha) \int \d\vr \ \bar{n}(\vr) w^2(\vr) \L_\ell (\vkhat \cdot \vrhat), 
\end{equation}
and we assume that $P_\ell^{\rm noise} = 0$ for $\ell > 0$, as it is 
negligible relative to $\widehat{P}_\ell$. 
We then approximate the line-of-sight to the pair of objects with the line-of-sight
to a single pair member, as 
$\L_\ell(\vkhat \cdot \vrhat_h) 
\simeq \L_\ell(\vkhat \cdot \vrhat_2)$. This approximation renders 
the integrals in equation~\ref{eq:Pell-full} separable, yielding the so-called ``Yamamoto estimator''
 \cite{YamamotoEtAl06,BeutlerEtAl14}

\begin{equation}\label{eq:Pell-yama}
\widehat{P}_\ell^{\rm yama} = \frac{2\ell+1}{I} \int \frac{\d\Omega_k}{4\pi} 
									\left [ 
									\int \d\vrone \ F(\vrone) e^{i \vk \cdot \vrone}
                                    \int \d\vrtwo \ F(\vrtwo) 
                                    e^{-i \vk \cdot \vrtwo}
                                    \L_\ell(\vkhat \cdot \vrhat_2)
                                    - P_\ell^{\rm noise}(\vk)
                                    \right ].
\end{equation}
This approximate line-of-sight remains reasonably accurate over the typical
range of scales considered in wide-field galaxy surveys, although it will eventually 
break down on very large scales 
\cite{YooSeljak15,SamushiaBranchiniPercival15,SlepianEisenstein16}.

Recently, \cite{BianchiGil-MarinEtAl15} and \cite{Scoccimarro15} presented similar
algorithms to accelerate the evaluation of equation~\ref{eq:Pell-yama} 
for the monopole, quadrupole, and hexadecapole ($\ell=0,2,4$) using FFTs.
By decomposing the dot product $\vkhat \cdot \vrhat$ into its Cartesian components, 
they show that equation~\ref{eq:Pell-yama} for a given $\ell$ can be expressed  
as a sum over the Fourier transforms of the density field $F(\vr)$ weighted by
products of Cartesian vectors. The $N \mathrm{log} N $ scaling of the FFT
algorithm allows speed-ups of several orders of magnitude as compared to the 
naive summation implementation of equation~\ref{eq:Pell-yama}. 
The implementation of \cite{BianchiGil-MarinEtAl15,Scoccimarro15} requires 
$(\ell+1)(\ell+2)/2$ FFTs to evaluate each $\hat{P}_\ell$, 
meaning  $1 + 6 + 15 = 22$ FFTs for $\ell=0,2$, and $4$.

Rather than using a Cartesian decomposition, we use the spherical 
harmonic addition theorem (e.g. \cite{AWH13}, equation (16.57)) to factor the 
Legendre polynomial into a product of 
spherical harmonics each depending on only a single unit vector: 

\begin{equation}\label{eq:add-theorem}
\mathcal{L}_\ell(\vrhat_1 \cdot \vrhat_2) = \frac{4\pi}{2\ell+1} 
					\sum_{m=-\ell}^{\ell} \Ylm(\vrhat_1) \Ylm^\star(\vrhat_2).
\end{equation}
This approach has recently been used by \cite{SlepianEisenstein16} to accelerate
measuring the anisotropic 2PCF with the single-pair-member LOS estimator, as 
well as to accelerate the measurement of the 3PCF both with direct 
evaluations of the spherical harmonics \cite{Slepian15_direct3pt_alg} and using FFTs \cite{SlepianEisenstein16}. 
\cite{aniso3pcf} further explores the use of spherical harmonics for the anisotropic 3PCF.  

Using equation~\ref{eq:add-theorem}, the multipole estimator becomes

\begin{equation}\label{eq:Pell-ours}
\widehat{P}_\ell(k) = \frac{2\ell+1}{I} \int \frac{\d\Omega_k}{4\pi} F_0(\vk) F_\ell(-\vk),
\end{equation}
with

\begin{align}\label{eq:Fell}
F_\ell(\vk) & \equiv \int \d\vr \ F(\vr) e^{i \vk \cdot \vr} 
								\L_\ell(\vkhat \cdot \vrhat), \nonumber \\
			& = \frac{4\pi}{2\ell+1} \sum_{m=-\ell}^{\ell} 
            				\Ylm(\vkhat) \int \d\vr \ F(\vr) \Ylm^*(\vrhat) e^{i \vk \cdot \vr}.
\end{align}
The sum over $m$ in 
equation~\ref{eq:Fell} contains $2 \ell + 1$ terms, each of which can be computed 
using a FFT. Similar to \cite{BianchiGil-MarinEtAl15} and \cite{Scoccimarro15}, 
we find that the multipole moments can be expressed as a sum of Fourier transforms
of the weighted density field. The critical difference, however, is 
that by expanding the Legendre polynomial 
in terms of the orthonormal spherical harmonic basis we avoid redundant 
terms entering the summation for each multipole. For the purposes of memory efficiency, 
we evaluate equation~\ref{eq:Fell} using a real-to-complex FFT and 
use the real form of the spherical harmonics, given by

\begin{equation}\label{eq:real-Ylm}
    \Ylm(\theta, \phi) \equiv
    \begin{cases}
        \sqrt{\cfrac{2\ell+1}{2\pi}\cfrac{(\ell-m)!}{(\ell+m)!}}
        \,\L_{\ell}^m(\cos\theta)\cos m\phi & \quad m>0 \\
        \sqrt{\cfrac{2\ell+1}{4\pi}} \,\L_{\ell}^m(\cos\theta) & \quad m=0  \\
        \sqrt{\cfrac{2\ell+1}{2\pi}\cfrac{(\ell-|m|)!}{(\ell+|m|)!}}
        \,\L_{\ell}^{|m|}(\cos\theta)\sin|m|\phi & \quad m<0,
    \end{cases}
\end{equation}
where $\L_\ell^m$ is the associated Legendre polynomial. The spherical harmonics
can be expressed in terms of Cartesian vectors using equation~\ref{eq:real-Ylm}
and the usual relations to transform from spherical to Cartesian coordinates.
Thus, equations \ref{eq:Pell-ours} and ~\ref{eq:Fell}, combined 
with the spherical harmonic expressions in equation~\ref{eq:real-Ylm}, enable computation of the multipole moments of the density field for arbitrary $\ell$.

To compute each multipole, our implementation requires only $2 \ell + 1$ FFTs,
as compared to $(\ell+1)(\ell+2)/2$ when using the Cartesian 
decomposition of \cite{BianchiGil-MarinEtAl15,Scoccimarro15}. Often, we are 
concerned with computing all even-$\ell$ multipoles up to a given $\lmax$.
For this case, our implementation requires a total of 
$(\lmax+2)(\lmax+1)/2 \sim \mathcal{O}(\lmax^2)$ FFTs, as compared to 
the total of $(\lmax+2)(\lmax+4)(2\lmax+3)/24 \sim \mathcal{O}(\lmax^3)$ 
for the Cartesian expansion. For example, for 9 multipoles ($\lmax = 16$), 
our approach offers a factor of $525 / 153 \simeq 3.4$ improvement. 

\subsection{Wedges}\label{sec:wedges}

The power spectrum can be expressed in terms of the multipole basis used in section~\ref{sec:multipoles} as

\begin{equation}\label{eq:Ptrue}
P(k,\mu) = \sum_{\ell=0}^\infty P_\ell(k) \L_\ell(\mu),
\end{equation}
where the power spectrum is parametrized by 
the amplitude $k$ and the cosine
of the angle to the line-of-sight $\mu$. In linear theory \cite{Kaiser87}, 
only the $\ell=0,2,4$ multipoles contribute to the sum in 
equation~\ref{eq:Ptrue}, but nonlinear evolution generates non-zero moments for 
multipoles with $\ell > 4$, albeit with diminishing importance as 
$\ell$ increases. In practice, we must truncate the sum in 
equation~\ref{eq:Ptrue} at some $\lmax$. Thus, we define our 
estimator for clustering wedges, averaged over discrete $k$ and $\mu$ bins, as 

\begin{equation}\label{eq:wedge-est}
    \widehat{P}(k_i, \mu_m) \equiv \sum_{\ell=0}^{\lmax} \widehat{P}_\ell(k) \Lbar_{\ell}(\mu_m, \mu_{m+1}),
\end{equation}
where the multipole estimator $\widehat{P}_\ell$ can be evaluated using the 
implementation described in the previous section, and we have defined the 
mean Legendre polynomial across a wedge ranging from $\mu_m$ to $\mu_{m+1}$ as

\begin{equation}\label{eq:Lbar}
 \Lbar_{\ell}(\mu_m, \mu_{m+1}) = \frac{1}{\mu_{m+1}-\mu_m} \int_{\mu_m}^{\mu_{m+1}} 
 				\d\mu \; \L_{\ell}(\mu).
\end{equation}
Here and throughout this paper, hat denotes an estimator and 
subscripted $k$ and $\mu$ indicate binned quantities. We assume uniform wavenumber
bins and use $k_i$ to denote the center of the $i^{\rm th}$ bin. We allow
for non-uniform bins in $\mu$, labeling the $m^{\rm th}$ wedge with $\mu_m$ to
denote a bin ranging from $[\mu_m, \mu_{m+1}]$. 

\subsection{Implementation}
\label{sec:implementation}

We implement the multipole and wedge estimators as presented in 
sections~\ref{sec:multipoles} and \ref{sec:wedges} as part of the 
publicly available software toolkit 
\texttt{nbodykit}
\cite{nbodykit}.\footnote{https://github.com/bccp/nbodykit}
Our implementation is fully parallelized with
Message Passing Interface (MPI) and uses a Python
binding \cite{pfft-python} of the \texttt{PFFT} software by \cite{pippig2013pfft} 
to compute FFTs in parallel. We use the symbolic manipulation functionality 
available in the \texttt{sympy} Python package \cite{sympy} to compute the
spherical harmonic expressions in equation~\ref{eq:real-Ylm} in terms
of Cartesian vectors. This allows the user to specify the desired multipoles
at runtime, enabling the code to be used to compute multipoles of arbitrary $\ell$. 
Testing and development of the code was performed on the 
Cray XC-40 system Cori at the National Energy Research Supercomputing Center (NERSC), 
and the code exhibits strong scaling, with a roughly linear reduction in wall-clock
time as the number of available processors increases. When computing all even
multipoles up to $\lmax = 16$ (requiring in total 153 FFTs), our 
implementation takes roughly 90 seconds using 64 processors on Cori.

For the results presented in this work, we place the galaxies and random objects
on a Cartesian grid using the Triangular Shaped Cloud (TSC) prescription
to compute the density field $F(\vr)$ of equation~\ref{eq:FKP-density}. We use the
interlaced grid technique of \cite{SefusattiEtAl16} to limit the effects of
aliasing, and we correct for any artifacts of the TSC gridding using the 
correction factor of \cite{Jing05}. The interlacing scheme
allows computation of the FFTs on a $512^3$ grid with accuracy comparable 
to the results when using a $1024^3$ grid, but with a wall-clock time that is 
$\sim8\times$ smaller. When using interlacing, the catalog of galaxies is 
interpolated on to two meshes separated by half of the size of a grid cell. 
We sum these two density fields in Fourier space and inverse Fourier Transform 
back to configuration space. We then apply the spherical harmonic weightings of 
equation~\ref{eq:real-Ylm} to this combined density field 
and proceed with computing the terms in equation~\ref{eq:Fell}. 
The speed-up provided by interlacing is particularly 
powerful when computing large $\ell$ multipoles. When combined with TSC 
interpolation, we are able to measure power spectra up to the 
Nyquist frequency at $k \simeq 0.4 \ h \mathrm{Mpc}^{-1}$ with 
fractional errors at the level of $10^{-3}$ \cite{SefusattiEtAl16}. 

\section{Isolating transverse $\mu=0$ systematics}
\label{sec:systematics}

As discussed above,
cosmological information in the linear regime is limited to $\ell_{\rm max}=4$, 
so one may question the value of algorithms that go to $\ell_{\rm max}>4$. One 
reason is that in the nonlinear regime higher-order multipoles are generated, and 
their information can be used to constrain nonlinear RSD models. Another motivation
is measurement contamination from systematics that are predominantly 
localized to some part of a clustering wedge. 
In this section, we present a method to isolate and potentially 
remove systematics from our clustering estimators, assuming that the systematic
signal is dominant in the plane of the sky (i.e., angular), which is a common
issue for galaxy surveys.
The contamination in this case is confined
to predominantly transverse $\mu=0$ modes. We consider a toy model for 
the process of fiber assignment, a common issue for galaxy surveys where the physical
process of assigning galaxy targets to spectrograph fibers leads to incomplete
target selection and creates a systematic signal that must be accounted for. 
Our discussion is particularly relevant for the 
Dark Energy Spectrograph Instrument (DESI; \cite{LeviEtAl13}), 
as the process of fiber assignment has recently been shown 
in \cite{PinolEtAl17} to introduce a largely transverse systematic signal. 

\subsection{A toy model for fiber assignment}
\label{sec:toy-model}

We model the effect of a plane-of-the-sky systematic by suppressing the 
observed power spectrum by a Dirac delta function at $\mu=0$, as

\begin{equation}\label{eq:toy-model}
\Pobs(k,\mu) = P(k,\mu) - P_c(k) \dirac(\mu),
\end{equation}
where $\dirac$ denotes a one-dimensional Dirac delta function, and $P_c(k)$ is 
the power spectrum of the contamination signal and describes the amplitude of 
the clustering suppression.  Here, $P(k,\mu)$ is the true anisotropic 
power spectrum in the absence of systematics. In purely linear theory, 
$P(k,\mu)$ would be fully described by its $\ell=0, 2,$ and $4$ multipoles 
\cite{Kaiser87}.

\begin{figure}
\centering
\includegraphics[width=0.6\textwidth]{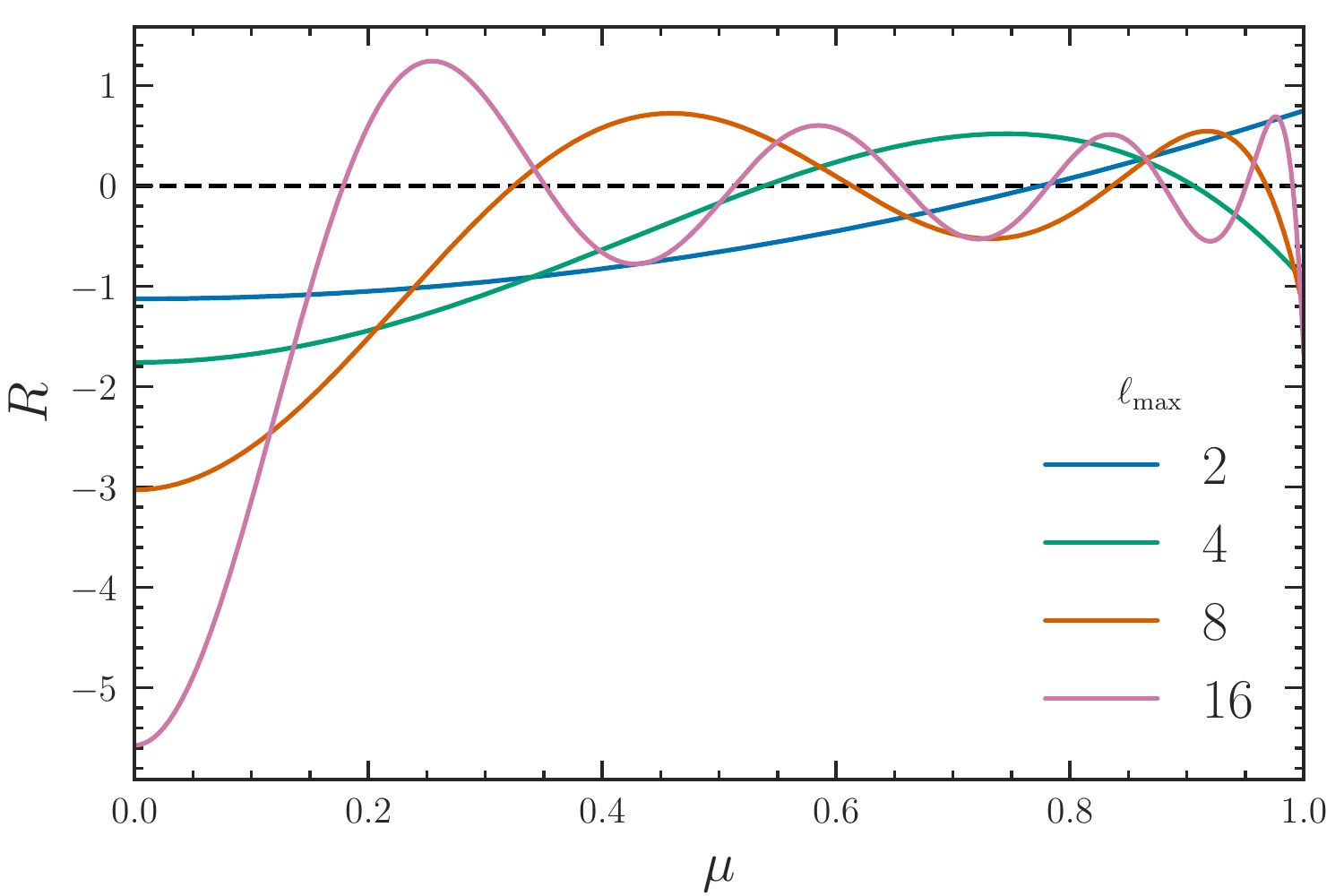}
\caption{The leakage of a transverse $\mu=0$ systematic into $\mu > 0$ power 
as a function of the maximum multipole used to reconstruct the observed
power $P(k,\mu)$. We plot the response of this error, as given in
equation~\ref{eq:response}. As multipoles are measured
to larger $\lmax$, the contamination is better isolated around
the origin, $\mu=0$.}
\label{fig:response}
\end{figure}

The contamination signal is localized in $\mu$ but affects all observed multipoles,
evident from the Legendre expansion of the Dirac delta function, 

\begin{equation}\label{eq:delta-ell}
\delta_{\ell} = \frac{2\ell+1}{2} \int_{-1}^{1} \d\mu \ \mathcal{L}_\ell(\mu) \dirac(\mu)
			=\frac{2\ell+1}{2} \mathcal{L}_\ell(0).
\end{equation}
In practice, we use only a finite number of multipoles, up to a desired $\lmax$, to
reconstruct the two-dimensional power spectrum $P(k,\mu)$. We can define
an estimator for the true power spectrum in the presence of a transverse systematic 
as 

\begin{align}\label{eq:Pobs-model}
\widehat{P}(k,\mu) & = \widehat{P}^{\rm obs}(k,\mu)
					+ P_c(k) \sum_{\ell=0}^{\lmax} 
                    	\frac{2\ell+1}{2}\L_\l(0)\L_\l(\mu), \nonumber \\
   & = \sum_{\ell=0}^{\lmax} \widehat{P}_\l(k) \L_\l(\mu) 
			+ P_c(k) \frac{\lmax+1}{2}\L_{\lmax+1}(0) \frac{\L_{\lmax+1}(\mu)}{\mu},
\end{align}
where our estimator for the observed power $\widehat{P}^{\rm obs}$ uses 
the measured multipoles $\widehat{P}_\ell$ up to $\lmax$, and we 
have used the Christoffel summation formula (\cite{GR} equation 8.915.1),

\begin{equation}
\sum_{\ell=0}^{\lmax}(2\ell+1)\L_\ell(x) \L_\ell(y) = \frac{\lmax+1}{y-x} 
				\left[
                \L_{\lmax}(x) \L_{\lmax+1}(y) - \L_{\lmax}(y) \L_{\lmax+1}(x)
                \right],
\end{equation}
with $x=0$ and $y=\mu$. Equation~\ref{eq:Pobs-model}
demonstrates that the $\mu=0$ contamination leaks into $\mu > 0$ modes because
of the finite number of multipoles used to reconstruct $P(k,\mu)$ and that
the angular dependence of this leakage is characterized by $\L_{\lmax+1}(\mu) / \mu$. 
We can describe the response of this leakage to the systematic signal as 

\begin{equation}\label{eq:response}
R(\mu) \equiv \frac{\widehat{P}^{\rm obs}(k,\mu) - \widehat{P}(k,\mu)}{P_c(k)} 
			= -\frac{\lmax+1}{2\mu} \L_{\lmax}(0) \L_{\lmax+1}(\mu).
\end{equation}
We show this response for various $\lmax$ values in figure~\ref{fig:response}.
While there is minimal signal in large $\ell$ multipoles, we can see from this figure
that the utility of measuring higher-order multipoles is that it enables sharper 
reconstruction of the angular dependence of the contaminating signal. By
increasing $\lmax$ we are able to increasingly localize the contamination
around $\mu=0$, with a width scaling as $\lmax^{-1}$.

\begin{figure}
\centering
\includegraphics[width=0.8\textwidth]{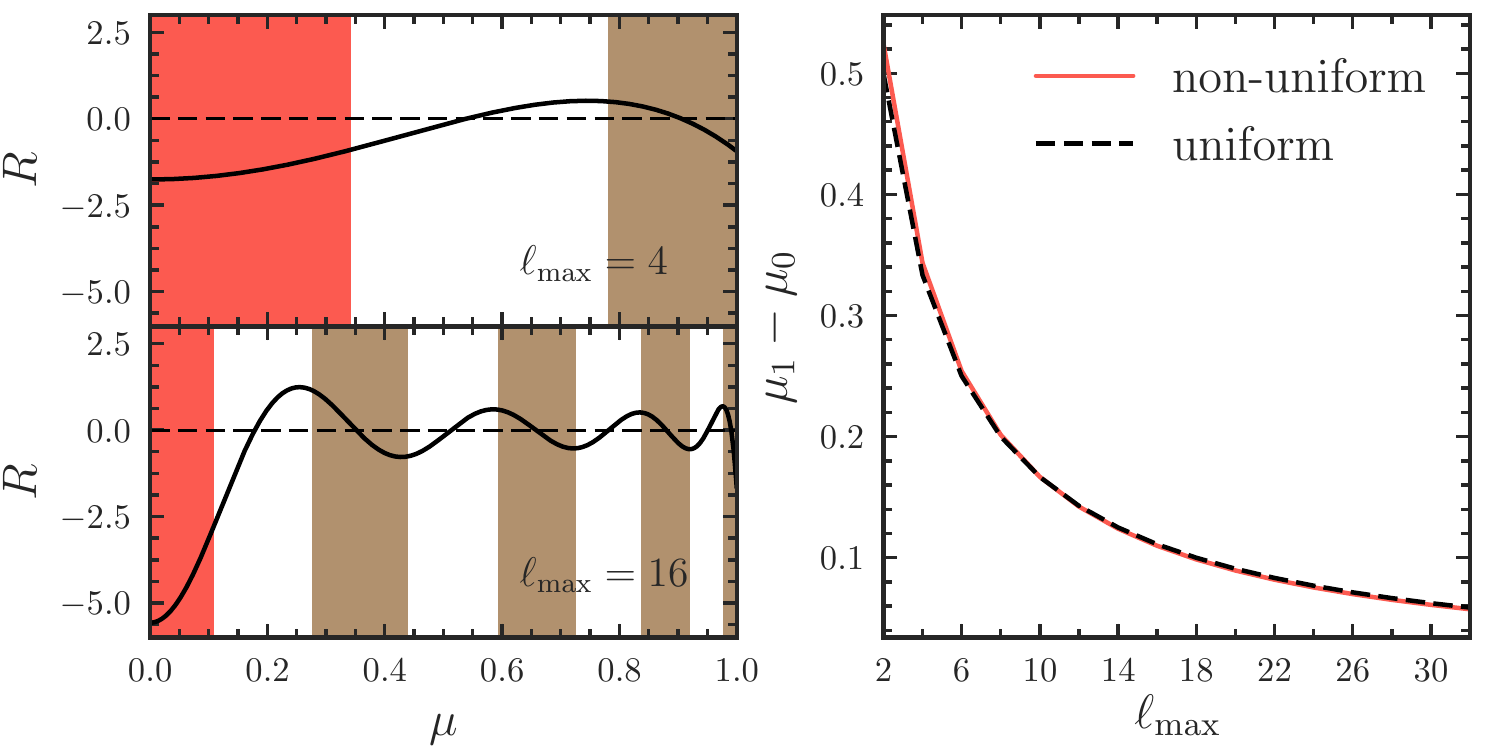}
\caption{Left: the leakage of a transverse $\mu=0$ systematic into $\mu>0$ power 
(black) for $\lmax = 4$ (top) and $\lmax = 16$ (bottom). We show the appropriate
non-uniform binning (shaded) that cancels the systematic in all but the 
first bin (red, shaded). Right: the width of the first $\mu\simeq0$ bin, given
by $\mu_1-\mu_0$, for the cases of non-uniform (red) and uniform (black, dashed) 
bins.}
\label{fig:nonuniform-bins}
\end{figure}

The oscillatory structure of the response in figure~\ref{fig:response} suggests
that we can employ a non-uniform binning in $\mu$ for our wedge estimator
of section~\ref{sec:wedges} in order to localize
the effect of the systematic to the first bin and 
cancel out the contamination in each of the other bins.  If we desire
to have as many wedge bins as number of observed multipoles (measuring even
multipoles up to $\lmax$), then there
will be $\lmax/2$ non-contaminated bins. The edges of these bins can 
be computed from the response in equation~\ref{eq:response} as

\begin{equation}\label{eq:nonuniform-bins}
    \int_{\mu_i}^{\mu_{i+1}} d \mu \ \frac{\L_{\lmax+1}(\mu)}{\mu}  \equiv 0,
    \quad i = 1, 2, \ldots, \lmax/2,
\end{equation}
where $\mu_i$ specifies the left edge of the $i^{\rm th}$ bin, and we have
assumed a total of $N_\mu = \lmax/2 + 1$ bins. By construction, we have
$\mu_0 = 0$ and $\mu_{\lmax/2+1} = 1$. In this notation, the only 
contaminated bin is the first, ranging from $\mu_0 < \mu < \mu_1$. 
We show the non-uniform binning for $\lmax=4$ and $\lmax=16$ as the 
shaded regions in the left panel of figure~\ref{fig:nonuniform-bins}. 
Generically, the $\mu$ wedges first become wider and then significantly
narrower ranging from $\mu=0$ to $\mu=1$.
We also show the width of the first, contaminated bin, $\mu_1-\mu_0$, 
in the right panel. The edge of the first bin closely
follows the result in the uniform case, $\mu_1-\mu_0 = (\lmax/2+1)^{-1}$. 
Larger $\lmax$ values clearly enable better isolation of the systematic signal in a
narrow first bin, and in turn, create a larger $\mu$ range absent
of any systematics.

\subsection{Verification with simulations}
\label{sec:sim}

\begin{figure}
\centering
\includegraphics[width=0.8\textwidth]{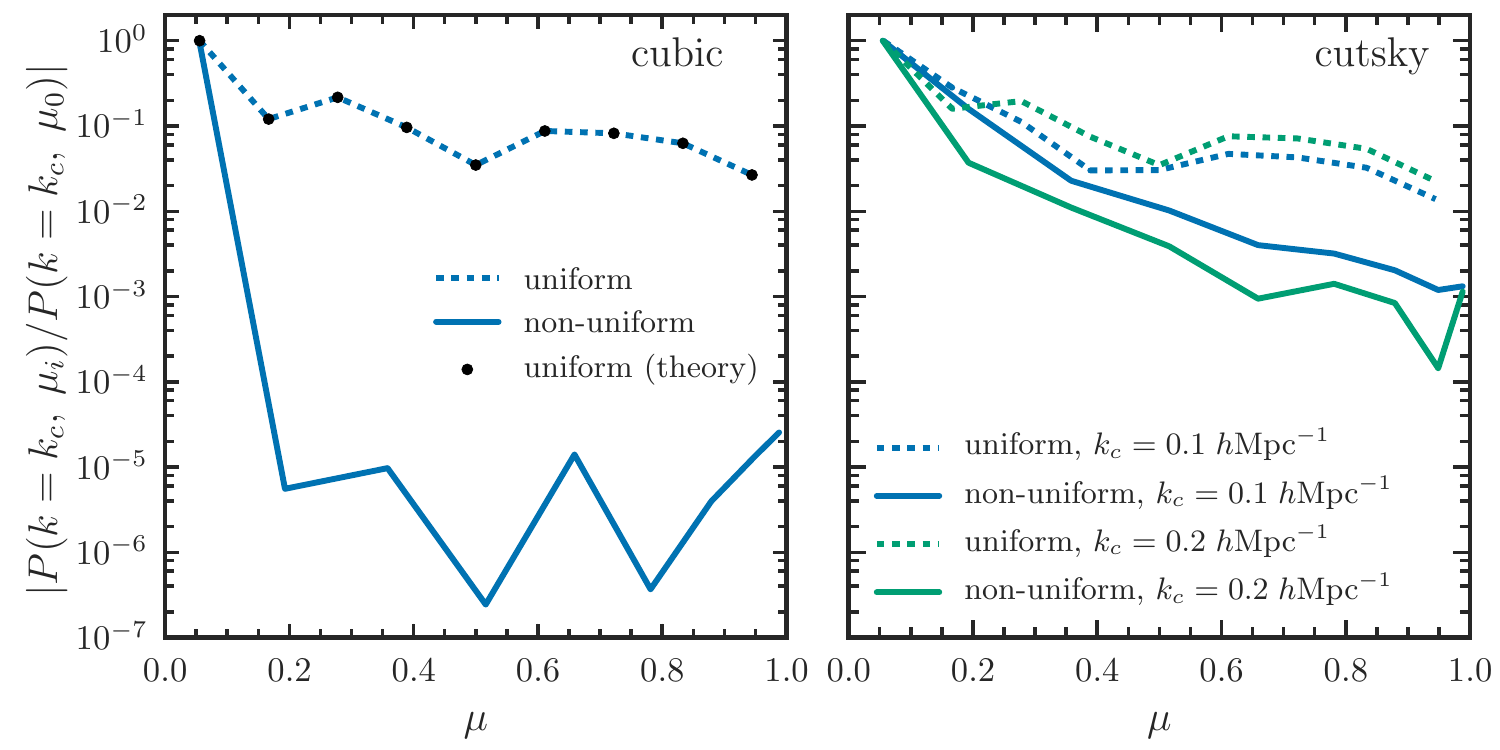}
\caption{The amplitude of a contaminating spike in 9 $P(k,\mu)$
wedges relative to its amplitude in the first $\mu$ bin for cubic simulation
boxes (left) and for cutsky mock catalogs with the BOSS DR12 selection function
imposed (right). Wedges are computed from even multipoles up to $\lmax=16$. 
The mock catalogs contain uniformly clustered objects with a
density field modulated via a sinusoidal function in the plane of the
sky, causing a large systematic spike in Fourier space at $k=k_c$. We show 
results for both uniform $\mu$ binning (dotted) and the non-uniform (solid)
scheme discussed in \S\ref{sec:toy-model}.}
\label{fig:sys-removal}
\end{figure}

We verify the utility of the non-uniform binning scheme
discussed in \S\ref{sec:toy-model} using simulated density fields. We generate
uniformly clustered catalogs of discrete objects and simulate an example
systematic signal by modulating the amplitude
of the density field in the plane of the sky. We use a sinusoidal function
for this modulation, which creates a large contaminating spike in 
Fourier space at a specific wavenumber, $k=k_c$. We perform this
test for both periodic boxes and for mock catalogs where the geometry
of the DR12 BOSS CMASS sample has been imposed \cite{AlamEtAl16,ReidEtAl16}.
We denote these latter mocks as cutsky mocks. For the cubic boxes, we
simply choose the $z$ axis of the box to be the line-of-sight and modulate
the amplitude of the density field in the ($x$,$y$) plane. For the cutsky 
mocks, which provide the angular and redshift coordinates of objects,
we apply the sinusoidal variation as a function of right ascension and
declination. We perform these tests for 50 cubic boxes of side length 
$L_{\rm box} = 2600 \; h^{-1}\mathrm{Mpc}$ and for 84 cutsky catalogs and compute
the average results to reduce noise.

We now compare our simulated results with the 
theoretical expectations from section~\ref{sec:toy-model}.
Because the catalogs are uniformly clustered, the true signal is a constant shot
noise that we can subtract from the results. We measure
the clustering wedges in both uniform and non-uniform bins and 
compare the amplitude of the contaminating spike at $k=k_c$ for each 
wedge relative to its amplitude in the first $\mu$ bin. The wedges 
are computed using even multipoles up 
to $\lmax=16$, which results in 9 $\mu$ wedges.
The left panel of figure~\ref{fig:sys-removal} shows the results for the cubic boxes.
We obtain near-perfect cancellation of the systematic when using 
non-uniform bins, isolating the contamination to only
the first bin at $\mu \simeq 0$. On the other hand, all wedges remain 
contaminated when using a uniform binning scheme. These results for uniform binning 
also agree well with our theoretical expectation (shown as black points), 
given the response in equation~\ref{eq:response}.

The removal of the systematic contamination using the cutsky catalogs, 
shown in the right panel of figure~\ref{fig:sys-removal}, is not as 
prominent as in the cubic case. However, the non-uniform binning
does reduce the amplitude of the systematic for all wedges, as 
compared to the uniform scheme, and this reduction is as large as
an order of magnitude for most bins. We perform two separate tests for 
the cutsky mocks, introducing systematic spikes at 
$k_c = 0.1 \; h\mathrm{Mpc}^{-1}$ and at 
$k_c = 0.2 \; h\mathrm{Mpc}^{-1}$. We find varying levels of success in eliminating
the systematic for these two cases, suggesting some unaccounted for $k$-dependence
in the optimal binning scheme. It is likely that the survey geometry, which is not 
present in the cubic case, complicates the simple model discussed 
in section~\ref{sec:toy-model}. In the cutsky case, the estimator measures
the power spectrum convolved with the window function. In particular, the 
systematic signal is also convolved with the window function, which 
mixes $k$ and $\mu$ modes and invalidates our simple
modeling assumptions in equation~\ref{eq:Pobs-model}. We expect the window function 
to be isotropic and have less influence on small scales (large $k$);
this is the trend we find
in our results, as we find better cancellation of the systematic
in the case of $k_c = 0.2 \; h\mathrm{Mpc}^{-1}$. We explore the effects
of the window function on our non-uniform binning scheme in more detail
in the next section. 

\subsection{A toy model for window function effects}
\label{sec:win}

Here, we outline a toy model to provide a qualitative understanding of the 
window function's impact on systematic removal. We show that the window 
function couples to the transverse systematic, effectively re-normalizing all of
its coefficients in the Legendre basis and thus implying a different choice of 
non-uniform bin boundaries relative to the window-free case for systematic elimination.  

We model the window function as a spherical top-hat in configuration space with 
radius $R$, so that

\begin{equation}
w(\vk; R) = \frac{3j_1(kR)}{kR},
\end{equation}
where $j_1$ is the spherical Bessel function of order one. The observed systematic 
is then convolved with the square of the window function as

\begin{equation}
P_c^{\rm win}(k,\mu) = \left\{w^2(k')\star P_c(k')\dirac(\mu) \right\}(\vk),
\end{equation}
where star denotes convolution. We note that $P_c^{\rm win}(k,\mu)$ remains a 
function only of $|\vk|$ and $\mu$ if the window function is isotropic, as in 
our toy model. We may evaluate this convolution using the Convolution Theorem, 
which gives

\begin{align}\label{eq:win-conv}
\left\{w^2(k')\star P_c(k')\dirac(\mu) \right\}(k,\mu) & = \nonumber \\
 & \FT\left\{\FT^{-1}\{w^2(k')\}(r)\;\FT^{-1}\{P_c(k')\dirac(\mu) \}(r) \right\}(k,\mu).
\end{align}
We first evaluate the inverse Fourier transform (FT) of $w^2(k)$. 
Applying the Convolution Theorem, the desired inverse FT is the convolution 
of two spherical top-hats, each of radius $R$ with centers separated by $r$.  
The overlap integral is given by the volume $V_{\rm lens}(r;R)$ of the spherical 
lens enclosed by both spheres when they are separated by $r$ \cite{Wolframlens},

\begin{equation}
V_{\rm lens}(r;R) = \frac{\pi}{12}(4R+r)(2R-r)^2.
\end{equation}
This result gives the first term inside the outer curly brackets in 
equation~\ref{eq:win-conv}. We now seek the second term, the inverse 
FT of the systematic.  Writing the Delta function using its Legendre 
expansion (equation~\ref{eq:delta-ell}) and then expanding the Legendre polynomials 
into spherical harmonics using the spherical harmonic addition theorem, we find

\begin{equation}
P_c(k')\dirac(\mu) = P_c(k')\sum_{\ell =0}^{\infty} \delta_{\ell} 
				\frac{4\pi}{2\ell+1}\sum_{m = -\ell}^{\ell} 
                		Y_{\ell m}(\vkhat')Y_{\ell m}^*(\vnhat),
\end{equation}
where $\delta_{\ell}$ is defined in equation~\ref{eq:delta-ell}.
The inverse FT can then be obtained by expanding the relevant exponential 
via the plane wave expansion into spherical Bessel functions and spherical 
harmonics (e.g., \cite{AWH13}, equation 16.52) and invoking orthogonality, leading to 
 
\begin{align}\label{eq:config-space-syst}
\FT^{-1} \left\{ P_c(k')\dirac(\mu) \right\} (r) = \sum_{\ell=0}^{\infty} 
 											S_{\ell}(r) \mathcal{L}_{\ell}(\mu),
\end{align}
where $\mu = \vrhat \cdot \mathbf{\hat{n}}$ and 
 
\begin{align}\label{eq:Sl-def}
S_{\ell}(r) = \int \frac{k'^2 \d k'}{2\pi^2}j_{\ell}(k'r) P_c(k').
\end{align}
We now have both terms in the outer curly brackets of 
equation~\ref{eq:win-conv} and simply require their product's Fourier transform 
to obtain $P_c^{\rm win}(k, \mu)$, the systematic observed in the presence 
of the window function. 

Expanding the Legendre polynomials of equation~\ref{eq:config-space-syst} 
into spherical harmonics using the addition theorem, again expanding the 
exponential via the plane wave expansion, and invoking orthogonality, 
we find

\begin{equation}\label{eq:Pc-win}
P_c^{\rm win}(k,\mu) = \sum_{\ell = 0}^{\infty}\mathcal{L}_{\ell}(\mu) \delta_\ell
					\int r^2 \d r\; V_{\rm lens}(r;R) S_{\ell}(r) j_{\ell}(kr).
\end{equation}
We pause to examine the limit where $R\to \infty$ and hence 
$V_{\rm lens}(r;R)$ is independent of $r$ and can be taken outside the 
integral; this corresponds to a boundary-free survey. In this limit, the 
integral over $r$ can be performed by substituting equation~\ref{eq:Sl-def}
and invoking the orthogonality relation for spherical Bessel functions and
we recover that $P_c^{\rm win}(k,\mu)\to P_c(k)\dirac(\mu)$.

We see that in general, in the presence of an isotropic window function, 
the coefficients of the Legendre expansion of $P_c(k,\mu)$ change and
are no longer given by the simple relation $\delta_{\ell} = (2\ell+1)/2 \; \L_\ell(0)$.
Importantly, they now have $k$-dependence, as the window function mixes 
the purely isotropic systematic amplitude $P_c(k)$ with the 
$\mu$-dependent Delta function. The non-uniform wedge boundaries of the 
previous section were set by the condition 
that for a given wedge, the sum of averaged Legendre polynomials weighted by 
the Delta function's coefficients would vanish. Here, we see that changing
these coefficients simply means this criterion is satisfied for a 
different non-uniform binning scheme.  

\begin{figure}
\centering
\includegraphics[width=0.6\textwidth]{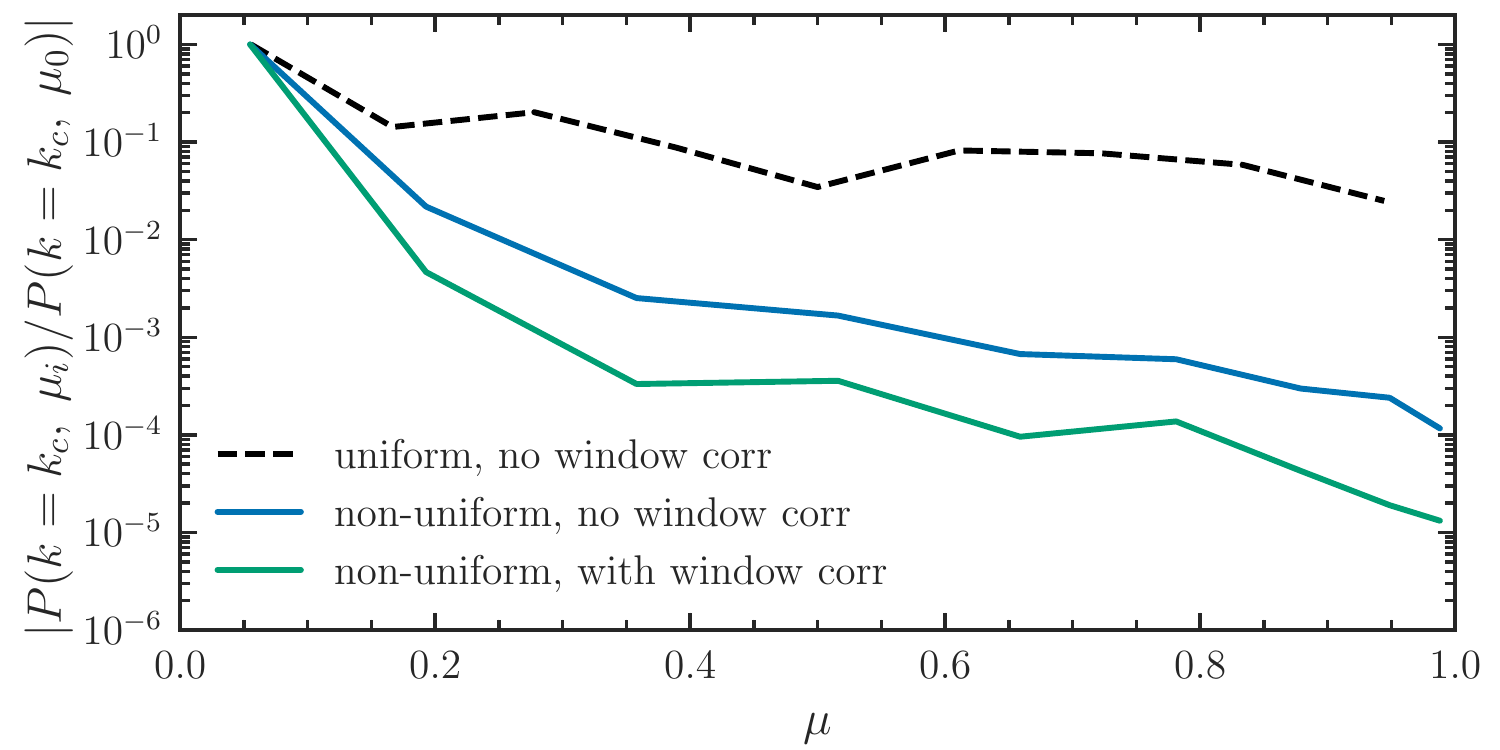}
\caption{The amplitude of a systematic spike in 9 $P(k,\mu)$ wedges
for a uniform clustering field with a toy-model selection function imposed
as described in \S\ref{sec:win}. When using a non-uniform binning scheme 
and accounting for window function effects, we can increase the success 
of the systematic removal by roughly an order of magnitude.}
\label{fig:windowed-sys-removal}
\end{figure}

We use simulations to examine the effectiveness of our non-uniform binning
scheme in the presence of this toy model window function. 
We apply a spherical top-hat window function of radius 
$R = 780 \; h^{-1}\mathrm{Mpc}$ to a uniformly clustered density 
field in a cubic box of side length 
$L_\mathrm{box} = 2600 \; h^{-1}\mathrm{Mpc}$. As in previous sections,
we model the systematic with a sinusoidal modulation of the density
field in the $(x,y)$ plane, assuming the $z$ axis represents the line-of-sight.
This modulation generates a spike in Fourier space at $k=k_c$, 
corresponding to the wavelength of the modulation. We once again
use the measured multipoles to estimate wedges in non-uniform bins, and
compare the results with and without accounting for the window function
corrections in equation~\ref{eq:Pc-win}. We present this comparison
in figure~\ref{fig:windowed-sys-removal}, which shows that by accounting
for the window function, an additional order of magnitude reduction
in the systematic signal can be achieved in all but the first $\mu$ wedge.
As in previous sections, we find that a uniform $\mu$ binning scheme
performs worse than our non-uniform scheme, even when ignoring window function
effects.

\section{Statistical properties}
\label{sec:stats}

In this section, we explore the covariance properties of the wedge 
estimator as a function of $\lmax$ and use a Fisher matrix formalism
to describe the effect on the derived parameter constraints when 
using our non-uniform binning approach to mitigate systematics.

\subsection{Covariance}
\label{sec:cov}

Under the assumption of purely Gaussian statistics, the covariance
of the power spectrum $P(k,\mu)$ averaged in bins of $k$ and $\mu$
is \cite{GriebEtAl16}

\begin{equation}\label{eq:full-wedge-cov}
\cov \bigl [ P(k_i, \mu_m), P(k_j, \mu_n) \bigr]
    = \delta_{ij} \delta_{mn} \frac{2}{N_{k_i} \Delta \mu_m}
    \!\int\! \frac{4\pi k^2 \d k}{V_{k_i}}\frac{\d \mu}{\Delta\mu_m}\,
    								\bigl[P(k, \mu) + \bar{n}^{-1} \bigr]^2,
\end{equation}
where the number density of the sample considered is $\bar{n}$, 
the volume of the shell in $k$-space is 
$V_{k_i} = 4\pi[(k_i + \Delta k/2)^3 - (k_i - \Delta k/2)^3]/3$, and the
number of modes in the $i^{\rm th}$ $k$ bin is 
$N_{k_i} = 4\pi k_i^2 \Delta k V_s / (2\pi)^3$, where $V_s$ is the volume
of the sample considered. Under the assumption of Gaussian statistics, different 
clustering wedges are not correlated, as reflected by the Kronecker 
delta factor $\delta_{mn}$ in equation~\ref{eq:full-wedge-cov}. 

The computationally-efficient estimator presented in this work does not 
directly measure the quantity $P(k_i, \mu_m)$ that enters into 
equation~\ref{eq:full-wedge-cov}. Rather, we reconstruct 
power spectrum wedges from a finite set of measured multipoles, up
to a specified $\lmax$. Thus, the relevant quantity is the covariance of the
multipoles averaged in $k$ bins, which is given by

\begin{equation}\label{eq:pole-cov}
    \cov \bigl [\widehat{P}_{\l}(k_i), \widehat{P}_{\lprime}(k_j) \bigr]
    = \delta_{ij} (2\l+1)(2\lprime+1) \frac{2}{N_{k_i}} 
    \!\int\! \d\mu \frac{2\pi k^2\d k}{V_{k_i}}\, \L_{\l}(\mu)\L_{\lprime}(\mu)
    											\bigl[P(k, \mu) + \bar{n}^{-1} \bigr]^2,
\end{equation}
where we see multipoles of different $\ell$ are correlated. From this 
covariance we can compute the covariance of the wedge estimator 
in equation~\ref{eq:wedge-est} as

\begin{equation}\label{eq:wedge-cov}
\cov\bigl[\widehat{P}(k_i, \mu_m), \widehat{P}(k_j, \mu_n)\bigr]
    = \sum_{\l=0}^{\lmax} \sum_{\lprime=0}^{\lmax} \Lbar_{\l}(\mu_m) \Lbar_{\lprime}(\mu_n)
    \cov\bigl[\widehat P_{\l}(k_i), \widehat P_{\lprime}(k_j) \bigr],
\end{equation}
where the mean Legendre polynomial across a wedge is given by equation~\ref{eq:Lbar}.

The wedge covariance in equation~\ref{eq:wedge-cov} is difficult to 
further simplify analytically, but before comparing
to simulations, we can make further progress using the simplifying assumption
of linear theory. In this case, we can use the Kaiser model \cite{Kaiser87}

\begin{equation}\label{eq:kaiser}
P(k,\mu) = (1+\beta\mu^2)^2 b_1^2 P_{\real}(k),
\end{equation}
where $\beta = f/b_1$ is the usual redshift-space distortion parameter, $b_1$
is the linear bias parameter, $P_{\real}(k)$ is the linear theory real-space power
spectrum, and $f$ is the logarithmic growth rate \cite{Kaiser87}.
Now, we can separate the scale and angular dependence in equation~\ref{eq:wedge-cov}.
We leave the scale dependence implicit in our notation to focus on the 
angular subspace in order to improve clarity. With these assumptions, 
the wedge covariance becomes

\begin{align}
\label{eq:trunc-Cmn}
    \widehat{C}_{mn} &\equiv
    \cov\bigl[\widehat{P}(\mu_m), \widehat{P}(\mu_n)\bigr]
    = \frac{2\gamma_{mn}}{N_{k_i}} \overline{P_{\real}^2},
\end{align}
where

\begin{equation}\label{eq:mean-Psq}
    \overline{P_{\real}^2} \equiv
    \int\!\frac{4\pi k^2\d k}{V_{k_i}}\, P_\real^2(k),
\end{equation}
and

\begin{equation}\label{eq:gamma}
    \gamma_{mn} \equiv
    \sum_{\l=0}^{\lmax} \sum_{\lprime0}^{\lmax} (2\l+1)(2\lprime+1)
    \Lbar_{\l}(\mu_m) \Lbar_{\lprime}(\mu_n)
    \int\!\frac{\d\mu}{2} \L_{\l}(\mu)\L_{\lprime}(\mu) (1+\beta\mu^2)^4.
\end{equation}
From these equations, we see that in the simple Kaiser model, the
correlation coefficient between between wedges $\mu_m$ and $\mu_n$,
defined as $\rho_{mn} = \widehat{C}_{mn} /  (\widehat{C}_{mm} \widehat{C}_{nn})^{1/2}$
is independent of scale with the amplitude proportional to the 
quantity $\gamma_{mn}$.

\begin{figure}[!tb]
\centering
\includegraphics[width=0.6\textwidth]{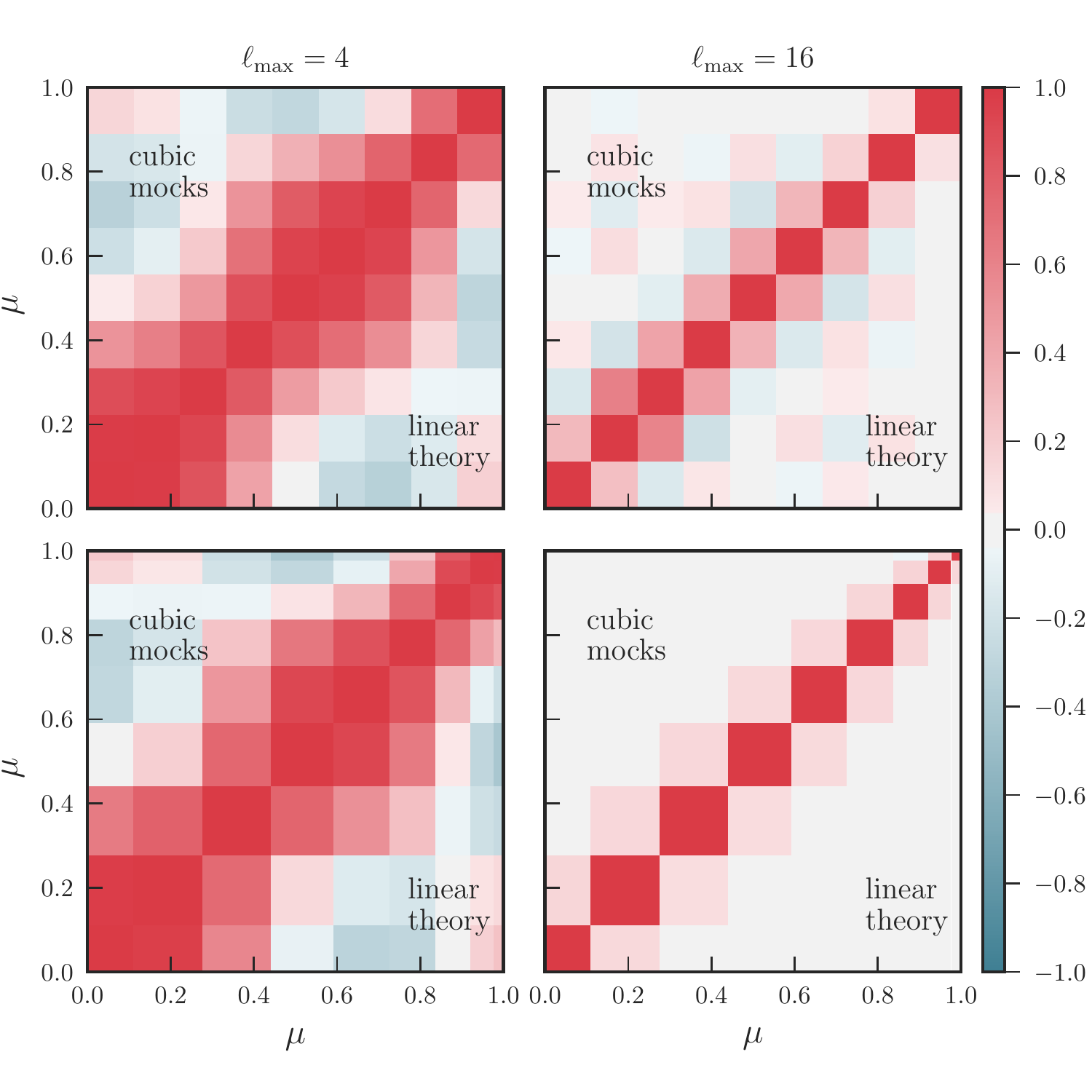}
\caption{The correlation matrix between $\mu$ wedges measured
from cubic box simulations (upper triangle) as compared to linear theory 
(lower triangle), when using uniform (top row) and non-uniform 
(bottom row) $\mu$ bins. For 9 $\mu$ wedges, we show results using
$\lmax=4$ (left column) and $\lmax=16$ to estimate the wedges from 
the corresponding multipoles. We find excellent agreement between linear theory
and the results measured from simulations.}
\label{fig:wedge-cov-cubic}
\end{figure}

\begin{figure}[!tb]
\centering
\includegraphics[width=0.6\textwidth]{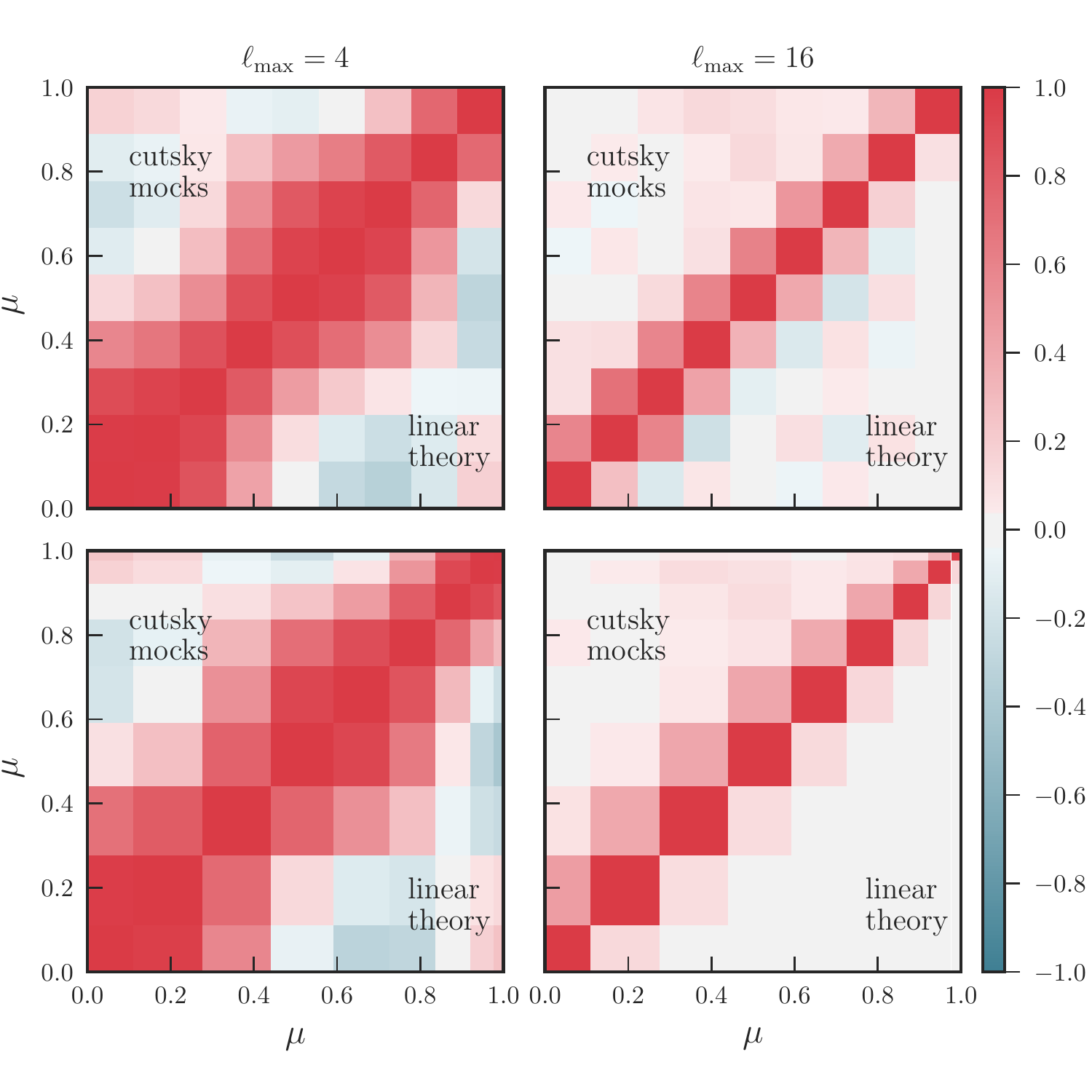}
\caption{The correlation matrix between $\mu$ wedges measured
from realistic cutsky mock catalogs (upper triangle) 
as compared to linear theory (lower triangle), when using uniform (top row) 
and non-uniform (bottom row) $\mu$ bins. For 9 $\mu$ wedges, we show results using
$\lmax=4$ (left column) and $\lmax=16$ to estimate the wedges from 
the corresponding multipoles. While there are discrepancies introduced by
the window function in comparison to the linear theory expectation,
the general trends remain consistent with the periodic box results. }
\label{fig:wedge-cov-cutsky}
\end{figure}

We first compare our simple theoretical modeling to the wedge covariance 
measured from 990 independent Quick Particle Mesh (QPM) 
periodic simulations \cite{White14b} at a redshift of $z=0.55$ and with a box size of 
$L_{\rm box} = 2560 \; h^{-1}\mathrm{Mpc}$. These simulations were designed
to mimic the clustering of the BOSS CMASS sample, with 
a linear bias of $b_1 \sim 2$ at $z \sim 0.5$.
We estimate the clustering wedges using the measured multipoles up to 
a specified $\lmax$ and use the 990 realizations to estimate the covariance 
of the wedges. We show the resulting correlation matrix between
separate $\mu$ wedges in figure~\ref{fig:wedge-cov-cubic} and compare
to the linear Kaiser result from equation~\ref{eq:gamma}. We perform
this comparison using both non-uniform (bottom row) and 
uniform (top row) binning schemes, as well as for $\lmax = 4$ (left column)
and $\lmax = 16$ (right column). In all cases, the number of wedges
is fixed to $N_\mu = 9$. We find excellent agreement between
a simple Kaiser model with $\beta = 0.35$ and the simulation
results. As expected, we find the wedges to be
significantly more correlated when using only three multipoles to reconstruct
nine $\mu$ wedges, as is the case for $\lmax=4$, than when using nine
measured multipoles, as for $\lmax = 16$. Furthermore, in the
case of $\lmax=16$, we find that our non-uniform binning scheme
achieves a significantly more diagonal covariance matrix between wedges, 
as seen in the right column of figure~\ref{fig:wedge-cov-cubic}. As the
matrix becomes more diagonal, the covariance is better approximated
by the Gaussian case, where the clustering wedges are fully independent.

We also compare our Kaiser modeling to a set of cutsky mock catalogs that 
include selection function effects, although we do not expect the 
simulation results to be well-described by this theoretical model in 
this case. We use a set of 84 mock catalogs which mimic the radial and 
angular selection functions of the BOSS DR12 CMASS sample 
\cite{ReidEtAl16,AlamEtAl16}. They model the 
true geometry, volume, and redshift distribution 
of the CMASS sample and were 
constructed from a set of seven independent, periodic box 
$N$-body simulations with the same cosmology and
a side length of $L_\mathrm{box} = 2600 \ h^{-1}\mathrm{Mpc}$.
Each of the 84 mock catalogs is an independent realization, and the clustering
of these cutsky catalogs is very similar overall to the 
BOSS CMASS sample at $z \sim 0.5$. As was done for the cubic box 
simulations, we compare simulation
and theory for the correlation matrix for nine $\mu$ wedges using 
$\lmax = 4$ and $\lmax = 16$. These results are presented in 
figure~\ref{fig:wedge-cov-cutsky}.
As expected, the cutsky simulation results are not as well-described
by the Kaiser model as in the cubic case due to window function effects. 
However, the general trends in the covariance are similar for the cutsky
case as for the cubic case. Importantly, we once again find that
using a higher $\lmax$ at fixed $N_\mu$ de-correlates the wedges and 
that the covariance is more diagonal when using our non-uniform binning scheme.


\begin{figure}[!tb]
\centering
\includegraphics[width=0.8\textwidth]{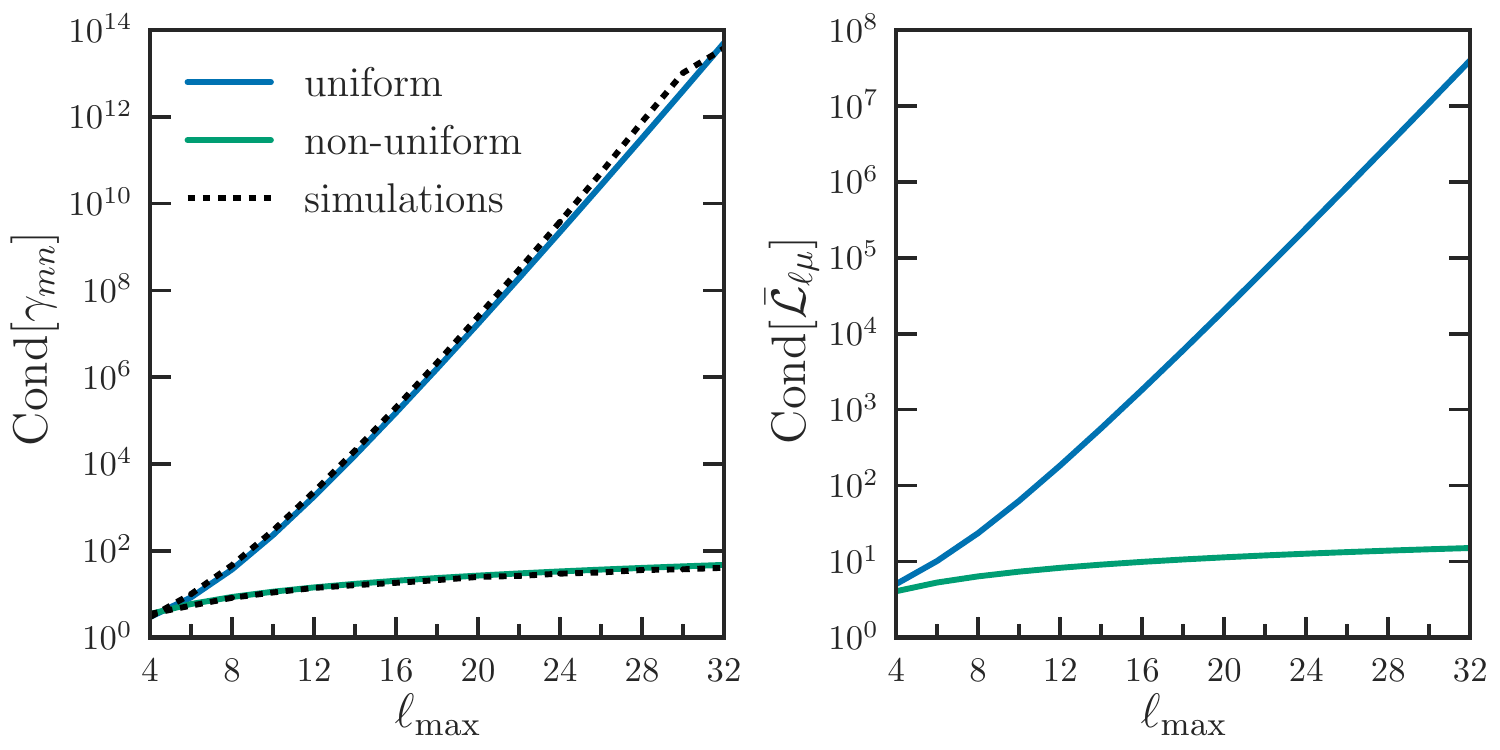}
\caption{Left: the condition number of the covariance matrix 
$\gamma_{mn}$, assuming a linear Kaiser model with $\beta=0.35$, 
as compared to the result computed from the 990 QPM simulations (black).
The simulation results have been re-normalized to match
the theoretical condition number at $\lmax = 4$. 
Right: the condition number of the matrix specifying
the mean Legendre polynomial across each $\mu$ wedge 
$\bar{L}_{\ell \mu}$ (right), as given by \eqref{eq:Lbar}.
We see that the wedge covariance matrix becomes ill-conditioned
for large $\lmax$ values when using a uniform binning scheme, 
driven by the fact that the transformation matrix $\Lbar_{\mu \ell}$ 
also becomes ill-conditioned.}
\label{fig:condition}
\end{figure}

An additional disadvantage of using bins uniform in $\mu$ is that the 
wedge covariance matrix quickly becomes ill-conditioned for high $\lmax$. 
This can be seen in figure~\ref{fig:condition}, where the left panel shows 
the condition number of the $\gamma_{mn}$ matrix as a function of 
$\lmax$ for both uniform and non-uniform bins. Here, the condition number 
of a matrix \textbf{M} is defined as the ratio of its smallest to largest singular 
values, computed using Singular Value Decomposition (SVD) 
(e.g., \cite{PressTeukolskyEtAl92}). The SVD of a matrix is defined 
as $\mathbf{M} = \mathbf{U}\mathbf{\Sigma} \mathbf{V}^T$, 
where $\mathbf{\Sigma}$ is a diagonal matrix with the 
singular values along the diagonal. 
We find similar trends for the condition number of the covariance matrix 
for our theoretical results assuming a linear Kaiser model with 
$\beta=0.35$ and for results computed from the 990 QPM boxes. 
Both uniform and non-uniform binning result in a reasonable condition 
number for $\lmax=4$, but the matrix in the uniform case becomes
increasingly singular as $\lmax$ increases. In such a case, the inversion
of the covariance, which is a necessary step of any likelihood analysis, 
becomes numerically unstable. This behavior at large $\lmax$ is largely
driven by the $\Lbar_{\ell\mu}$ matrix, which defines the contribution
of a multipole of order $\ell$ to a given $\mu$ wedge. The condition
number of this matrix is shown in the right panel of figure~\ref{fig:condition},
and its behavior mirrors that of the full covariance matrix. 

The functional form of $\L_\ell(\mu)$ can provide some insight into the large
condition number of the covariance matrix when using uniformly spaced bins. The 
Legendre polynomial of order $\ell$ oscillates around zero, and the frequency
of the oscillation increases with increasing $\mu$. For large $\lmax$, there
exist bins at $\mu \sim 1$ where the Legendre polynomial exhibits a 
positive/negative symmetry across the bin, and thus, the average value 
cancels very nearly to zero. This presents problems in 
equation~\ref{eq:wedge-est}, where our measured 
multipoles are weighted by the mean Legendre polynomial. These issues
are mitigated by our non-uniform bins, which were constructed such that 
the width of the bins decreases as a function of $\mu$, just as 
the Legendre polynomials oscillate more quickly. 
Thus, the bin cancellation is mostly avoided when non-uniform bins 
are used and the condition number of the resulting covariance
matrix remains stable, even at large $\lmax$. 
Such a binning scheme becomes appealing for clustering analyses, even 
if systematic mitigation is not the primary goal. 

\subsection{Fisher information}
\label{sec:info}

We can evaluate the information content of our wedge estimator as a function
of $\lmax$ using the Fisher matrix formalism. 
As in section~\ref{sec:cov}, we assume a simple linear Kaiser model
(equation~\ref{eq:kaiser}), where the parameter vector of interest is
$\mathbf{p} = (b_1 \sigma_8, f \sigma_8)$. For clarity, we also suppress the 
$k$ indexing here, as the $\mu$ and $k$ dependence of the covariance is fully
separable for the Kaiser model.
Assuming a Gaussian likelihood function for the clustering wedge observables, 
we can express the Fisher matrix as

\begin{equation}\label{eq:Fij-all}
F_{ij} = \sum_{m=0}^{N_\mu-1} \sum_{n=0}^{N_\mu-1} 
					\frac{\partial P(\mu_m)}{\partial p_i} \widehat{C}_{mn}^{-1} 
                    \frac{\partial P(\mu_n)}{\partial p_j},
\end{equation}
where $N_\mu$ is the number of (non-uniform) $\mu$ bins, $P(\mu_n)$ is the theoretical
Kaiser model averaged over the $\mu_n$ wedge, and the covariance between 
the measured wedges $\widehat{C}_{mn}$ is given by equation~\ref{eq:trunc-Cmn}.
We can also use this formalism to quantify the cost of removing the first $\mu$ bin 
when using our non-uniform binning scheme. In this case, the 
Fisher matrix is given by

\begin{equation}\label{eq:Fij-nomu0}
F^{\mu \not \simeq 0}_{ij} = \sum_{m=1}^{N_\mu-1} \sum_{n=1}^{N_\mu-1} 
					\frac{\partial P(\mu_m)}{\partial p_i} \widehat{C}_{mn}^{-1} 
                    \frac{\partial P(\mu_n)}{\partial p_j},
\end{equation}
where we have explicitly removed the contribution from the $\mu_0$ wedge
to the double sum in this equation. 

\begin{figure}[!tb]
\centering
\includegraphics[width=0.8\textwidth]{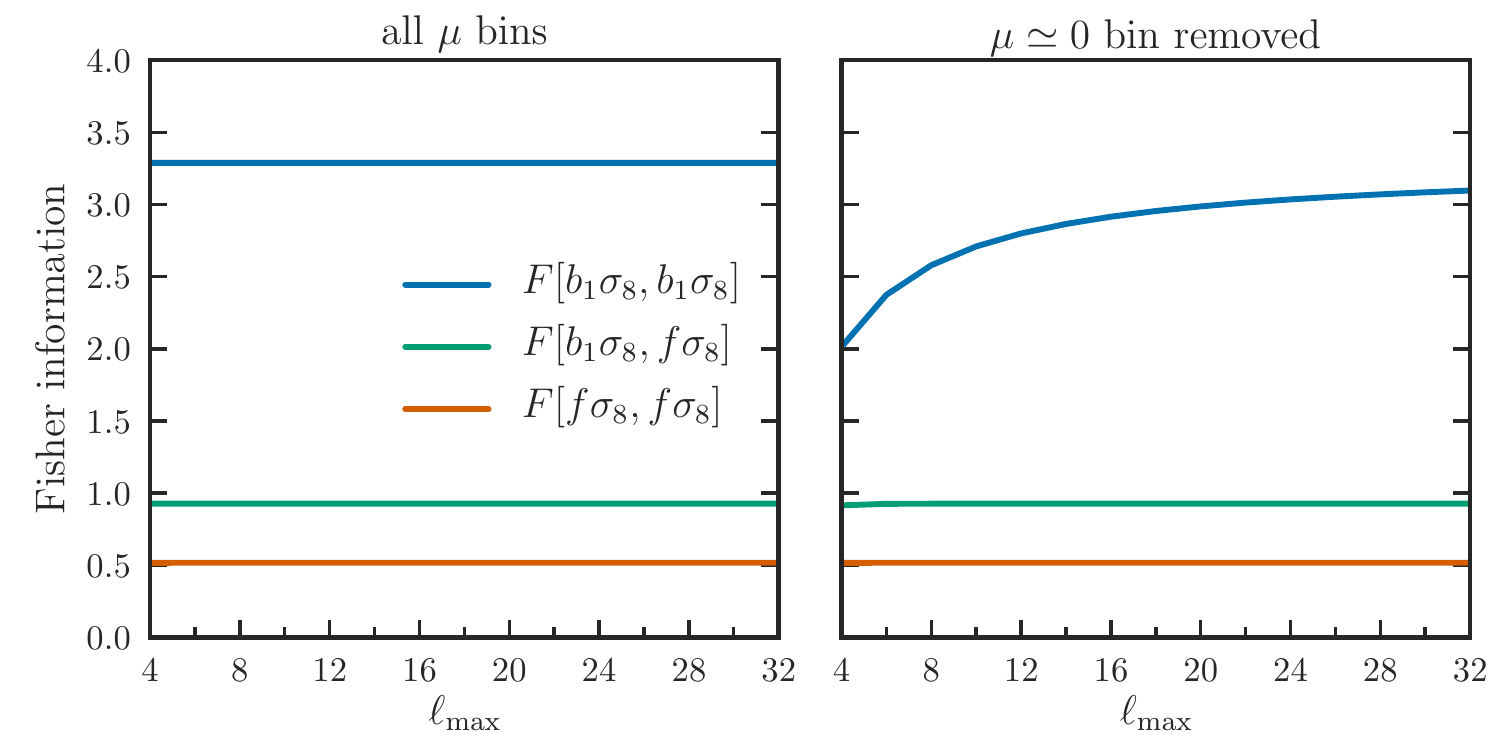}
\caption{The Fisher information for the parameter vector 
$\mathbf{p} = (b_1 \sigma_8, f \sigma_8)$ for our wedge estimator
using non-uniform $\mu$ bins, in the case of the 
linear theory Kaiser model. We show results as a function of the 
maximum multipole used to reconstruct the clustering wedges,
as well as the case when using all $\mu$ bins (left) and
when excluding the first $\mu$ bin (right).  A linear Kaiser model 
with $\beta=0.35$ has been assumed.}
\label{fig:fisher-info}
\end{figure}

We show the Fisher information for the auto-correlations of $b_1 \sigma_8$
and $f \sigma_8$, as well as their cross correlation, as a function of 
$\lmax$ in figure~\ref{fig:fisher-info}. Results are computed for the non-uniform
$\mu$ binning scheme presented in section~\ref{sec:toy-model}, assuming
a value of $\beta=f/b_1=0.35$ for the Kaiser model.
The left panel shows the information content when using all $\mu$ bins, 
and as expected, the information content saturates at $\lmax=4$ because only 
the $\ell=0,2,$ and $4$ multipoles are 
non-zero in the Kaiser model. In the right panel of this figure, we
show the Fisher information when we exclude the first $\mu$ bin from the analysis.
In this case, the information on $b_1 \sigma_8$ is partially lost, approximately 
proportional to the width of the missing wedge. However, the information on $f \sigma_8$
remains relatively unaffected by the missing wedge. The 
first wedge at $\mu \simeq 0$ is a prominent source of information on 
the amplitude of the power spectrum, as parametrized by $b_1\sigma_8$, 
but contains little information on the $\mu$ dependence of the clustering.

\begin{figure}[!tb]
\centering
\includegraphics[width=0.6\textwidth]{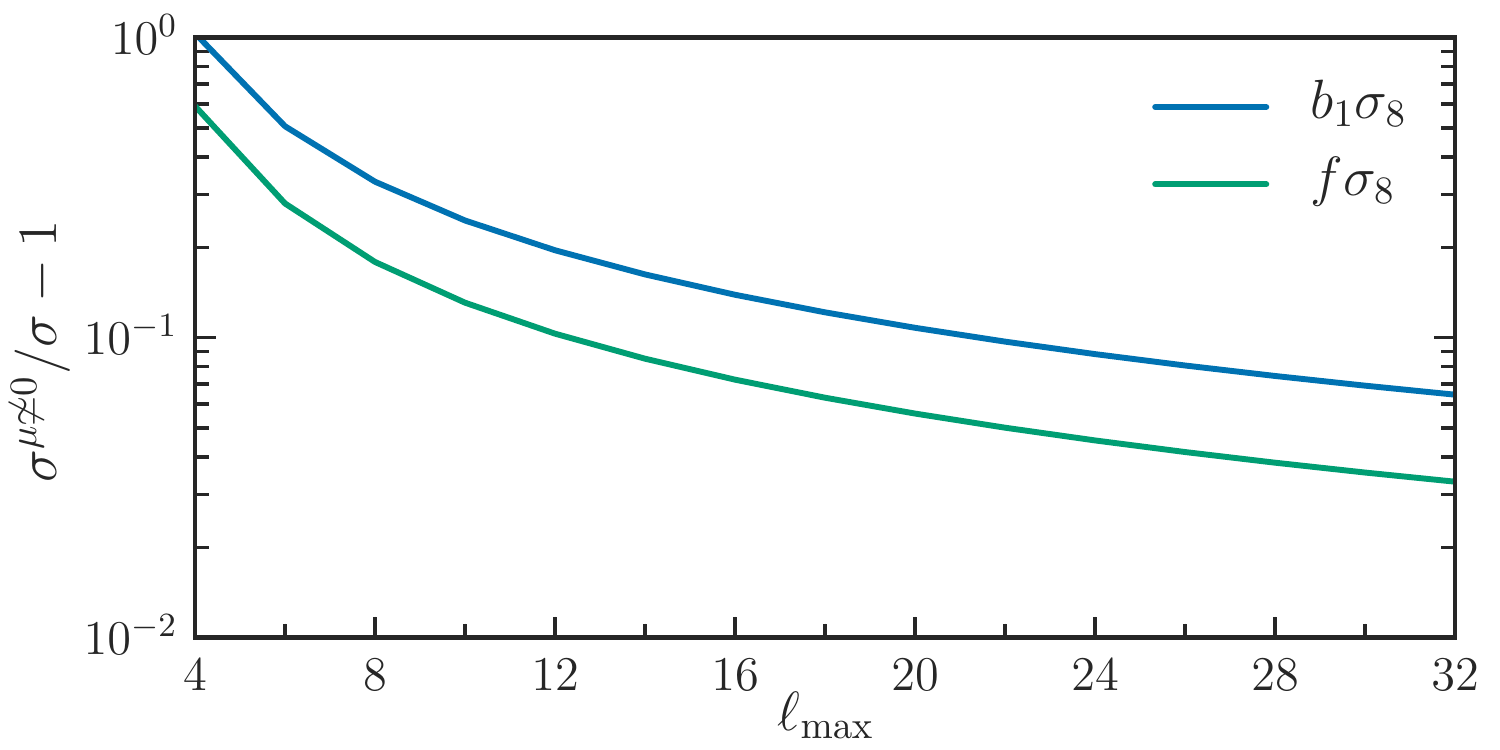}
\caption{The fractional change in the uncertainties in $b_1 \sigma_8$
and $f \sigma_8$ when using non-uniform $\mu$ wedges and
excluding the first $\mu \simeq 0$ wedge from the analysis, as determined 
from the Fisher matrix. A linear Kaiser model with $\beta=0.35$ 
has been assumed. }
\label{fig:fisher-errors}
\end{figure}

The inverse of the Fisher matrix provides an estimate of the marginalized 
error on a given parameter, such that the error on the parameter $A$ is given by
$\sigma_A = (F^{-1})_{AA}^{1/2}$. Thus, we can use the Fisher formalism to
evaluate the change in the parameter uncertainties when excluding 
the first $\mu\simeq0$ wedge in the presence of a transverse systematic. 
We show this fractional change for $b_1 \sigma_8$
and $f \sigma_8$ as a function of $\lmax$ in figure~\ref{fig:fisher-errors},
and we find the loss of constraining power drops rapidly with $\lmax$. For
$\lmax = 16$, we find $\sim7$\% and $\sim13$\% increases in the
uncertainties on $f \sigma_8$ and $b_1 \sigma_8$, respectively,
as compared to $\sim54$\% and $\sim92$\% for $\lmax=4$. With a reasonably 
large choice for $\lmax$, we can exclude the contaminated $\mu \simeq 0$ bin
with only marginal losses for the parameter constraints of interest. 

\section{Conclusions}
\label{sec:conclusions}

In this work, we have presented an optimal estimator for the anisotropic power 
spectrum multipoles that is valid in the local plane-parallel approximation. 
Our implementation eliminates redundancy present in previous algorithms 
\cite{BianchiGil-MarinEtAl15,Scoccimarro15}. These works rely on a 
Cartesian decomposition of the Legendre
basis to write the power spectrum estimator of \cite{YamamotoEtAl06}
using Fast Fourier Transforms. We improve upon them by 
using a spherical harmonic decomposition of the Legendre polynomials, 
motivated by the approach of \cite{SlepianEisenstein16} for the anisotropic 2PCF.
The method presented here is substantially faster than previous anisotropic 
power spectrum algorithms and renders calculation of multipoles to 
high $\lmax$ computationally feasible. For a given multipole of order 
$\ell$, our method requires only $2\ell+1$ FFTs rather than the 
$(\ell+1)(\ell+2)/2$ FFTs of the Cartesian approach. For the highest $\lmax$ 
used in this work, $\lmax = 16$, our approach is $\sim3.4\times$ faster 
than previous works, using 153 FFTs as opposed to 525.

Our estimator's significant reduction in wall-clock time allows construction 
of finely-binned wedges in $P(k,\mu)$ by combining multipoles up to 
high $\lmax$. We show that narrow $\mu$ bins
are particularly advantageous for mitigating the effects of systematic
contamination in the plane of the sky, as is often the case for galaxy 
surveys (e.g., \cite{PinolEtAl17}). In the presence of such an
angular systematic signal, we show that a non-uniform binning scheme in $\mu$ 
can effectively isolate the contamination to the first $\mu \simeq 0$ wedge 
and that the systematic contributions to all other bins can be eliminated.
We have verified the effectiveness of our non-uniform
bins on both periodic simulations and realistic mock catalogs that
have a survey selection function. We have demonstrated
with a toy model that a survey selection function mixes the $k$ 
and $\mu$ dependence of the systematic signal, introducing $k$-dependence
into the optimal non-uniform wedge boundaries. However, the systematic
signal can still be reduced even when ignoring these effects.
When analyzing galaxy survey data, knowledge of the window function 
and realistic simulations can be used to choose the optimal binning 
to reduce transverse systematics. 

We have also explored the statistical properties of the wedge estimator
as a function of the maximum measured multipole $\lmax$. We show using
linear theory that  
the covariance of the wedge estimator quickly becomes ill-conditioned for 
large $\lmax$ when using uniform bins, and we verify this finding with simulations. 
Consequently, when using uniform bins the covariance inversion is numerically 
unstable, creating a significant barrier for any likelihood analysis. On the other 
hand, the non-uniform binning scheme described in this work remains 
well-conditioned for all $\lmax$ values, enabling its inversion and use 
in model fitting. We also show
that at a fixed number of $\mu$ wedges, using larger values of $\lmax$
de-correlates separate wedges, and that the covariance matrix of
wedges using non-uniform bins is more diagonal than in the uniform case.
With a Fisher analysis assuming linear theory, we have demonstrated 
that the uncertainty on $f \sigma_8$ inflates by $\sim7$\% with $\lmax=16$ when excluding
the first $\mu$ wedge, assuming it is fully contaminated by systematics, 
as compared to a 54\% increase with $\ell_{\rm max}=4$. 
Even larger choices for $\lmax$ can further reduce this increase and should
be explored in more detail for future RSD analyses in the presence of
transverse (angular) systematics. 

We note that similar techniques as those presented in this paper can
be applied to clustering wedges in configuration space. However, the choice
of optimal non-uniform bins to remove systematics is further complicated 
for a correlation function analysis, 
as the systematic signal is no longer localized to $\mu=0$. Importantly, 
the optimal binning choice becomes a function of both the separation
perpendicular and parallel to the line-of-sight, $r_\perp$ and $r_\parallel$, 
which introduces additional modeling complexity. Similar techniques in configuration 
space should be further explored to assess their effectiveness at minimizing
the effects of angular systematics. 

Finally, we also point out that, as shown in \cite{Slepian15_wide_angle} for the anisotropic 2PCF, slight generalizations of the local plane parallel multipole estimates can be combined to yield the separation midpoint or angle bisector method-based multipoles. This point is important because it enables midpoint and bisector-based multipoles to be obtained by FFTs. As the relevant geometry for anisotropic clustering is the same in Fourier space and configuration space, combining \cite{Slepian15_wide_angle} with the results of this work will enable estimation of midpoint or bisector-based multipoles to very high $\lmax$ with FFTs, relevant for properly handling wide-angle effects in next-generation surveys.

The improvements to the power spectrum estimator 
presented in this work will prove valuable for
next generation redshift surveys such as DESI 
\cite{LeviEtAl13,AghamousaEtAl16a,AghamousaEtAl16b} and Euclid \cite{LaureijsEtAl11} both 
for the data measurement and for the covariance estimation, which requires 
analyzing a large number of mock catalogs. Given these surveys' large volumes 
and consequent high statistical precision, an unprecedented level of systematics 
control is required. The non-uniform clustering wedges described in this work 
will be important in this regard for DESI (recently described in \cite{PinolEtAl17}). 
In the future, these methods should be developed and further tested on realistic end-to-end simulations of upcoming surveys. 

\acknowledgments

NH is supported by the National Science Foundation Graduate Research Fellowship under 
grant number DGE-1106400. US is supported by NASA grant NNX15AL17G. ZS acknowledges support from a Chamberlain Fellowship at Lawrence Berkeley National Laboratory and from the Berkeley Center for Cosmological Physics.

\bibliographystyle{JHEP}
\bibliography{cosmo,cosmo_preprints,local}
\end{document}